\documentclass[twocolumn,showpacs,amsmath,amstex,amssymb,mathfonts,prl]{revtex4-1}
\usepackage{graphicx}

\usepackage{tipa}
\usepackage{bm}

\usepackage{amsmath}
\usepackage{amssymb}
\usepackage{amsthm}
\usepackage{amsfonts}

\begin{document}

\title{Cluster Statistics and Quasisoliton Dynamics in Microscopic Optimal Velocity Models}

\author{Bo Yang$^1$, Xihua Xu$^{2,3}$, John Z.F. Pang$^{1}$, and Christopher Monterola$^1$}
\affiliation{$^1$ Complex Systems Group, Institute of High Performance Computing, A*STAR, Singapore, 138632.}
\affiliation{$^2$ Department of Mathematics, National University of Singapore, 119076, Singapore.}
\affiliation{$^3$ Beijing Computational Science Research Center, Beijing 100084, PR China.} 
\date{\today}

\date{\today}
\begin{abstract}
Using the non-linear optimal velocity (OV) models as an example, we show that there exists an emergent intrinsic scale that characterises the interaction strength between multiple clusters appearing in the solutions of such models. The interaction characterises the dynamics of the localised quasisoliton structures given by the time derivative of the headways, and the intrinsic scale is analogous to the ``charge" of the quasisolitons, leading to non-trivial cluster statistics from the random perturbations to the initial steady states of uniform headways. The cluster statistics depend both on the quasisoliton charge and the density of the traffic. The intrinsic scale is also related to an emergent quantity that gives the extremum headways in the cluster formation, as well as the coexistence curve separating the absolute stable phase from the metastable phase. The relationship is qualitatively universal for general optimal velocity models.
\end{abstract}

\maketitle 

Modelling traffic flow, especially in an attempt to understand the occurrence of the traffic jams\cite{treiberbook,yukawa,yukawa1,rehborn,rehborn1,kerner1}, has been a fascinating subject leading to interesting development in many related fields. Several common approaches in modelling the evolution of the traffic flow include the microscopic car-following model\cite{bando,stepan,werner}, cellular automata\cite{schall,nagel} and the macroscopic hydrodynamic model\cite{seibold,kim}; more thorough reviews can be found in \cite{helbing, dogbe,as}. Most microscopic models involve anisotropic nearest neighbour interactions. One elegant class of models is the optimal velocity  (OV) models, with an explicit optimal velocity function dependent on the relative distance between the car and the next one ahead, or the headway\cite{bando}. Extensions of such models include additional force terms so that the acceleration or deceleration of the cars leaving/entering jammed region is not too large\cite{Helbing_PRE98,JiangR_PRE01, PengGH_PhyA13,GLW_PhyA08}. Other more realistic microscopic models include the intelligent driver models (IDM)\cite{helbingidm}, Shamoto's models\cite{shamoto} and various types of the sophisticated three-phase traffic models\cite{k1,k2,k3}. The relationships between these models are also explored in\cite{boyang}. Multiple preceding cars and even following cars are included to better model the driver decision-making process\cite{xue,hasebe}, and non-linear velocity difference effects are studied in\cite{xihua}. 

Controversies still remain on what aspects of real traffic dynamics can be captured by simple models like the optimal velocity model\cite{kerner2,schreck,helbing1}. Such models assume the existence of a fundamental diagram, thus all steady states have a unique relationship between the flow and density. This is in contrast with the fundamental assumptions of the three-phase traffic theories\cite{kernerbook}, that a multitude of steady states with non-unqiue flow-density relationship exists in the ``synchronised phase". While one does not expect such simple traffic models to capture all the empirical features of the congested traffic flow, these models offer a physically intuitive way to understand the formation of jams from the non-linear interactions between the system components, which are useful in designing intelligent mass transport systems\cite{hasebe} made of, for example, sensor equipped driverless cars. In addition, they have the potential to characterise a wide range of physical phenomena including the complex spatiotemporal traffic patterns, dynamics of (quasi-)one dimensional granular flows and the clustering of dissipative ``granular gases"\cite{zanetti}. It is thus of great theoretical interest to study the universal behaviours of these models especially in the non-linear regime.

In this paper, we do not concern ourselves with the capabilities of the models in capturing the empirical features of the traffic flow. Instead, we study the formal non-linear dynamics of the OV model class, especially focusing on the multi-cluster solutions. Using the original OV model as an example for its simplicity, we show that by properly non-dimensionalizing the model, the emergent symmetry of the cluster formation is rendered explicit, and the extremum headway of the clusters is an emergent quantity which gives the coexistence curve separating the absolutely stable and metastable phase of the model. Our numerical calculation shows that the probability distribution of cluster numbers depends both on an \emph{intrinsic scale} of the model and the density of the traffic lane. This can be explained by the dynamics of the ``quasisolitons" in the domain of headway velocity, which will be explained in details later. The strength of attraction between quasisolitons of opposite charges depends both on the intrinsic scale and the distance between them. The intrinsic scale is thus analogous to the charge of the quasisolitons.

A general car-following model can be written as
\footnotesize
\begin{eqnarray}\label{general}
\tau\dot v_n=-v_n+V\left(h_{n-i},\dot h_{n-i},\cdots,h_n,\dot h_n,\cdots,h_{n+j},\dot h_{n+j}\right)
\end{eqnarray}
\normalsize
where the \emph{dot} represents time derivative and $n\in \mathbb Z^+$ is the index of the cars; $v_n$ is the velocity of the $n^{\text{th}}$ car; $h_{n+i}$ is the distance between the $n^{\text{th}}$ car and the $(i+1)^{\text{th}}$ car in front of it, while $h_{n-i}$ is the distance between the $n^{\text{th}}$ car and the $i^{\text{th}}$ car behind it, which by convention is \emph{negative}. The first viscosity term on the right models the increasing tendency for the driver to decelerate when the car travels faster, and $\tau$ is the reaction time for the driver to maintain the optimal velocity given by the second term on the right. In this work the higher derivatives are suppressed as we assume the reaction time is small. The periodic boundary conditions gives $v_{N+n}=v_n$, where $N$ is the total number of cars. For physically relevant cases the optimal velocity is non-linear: it is generally assumed that $V$ is monotonically increasing for all its arguments, and it is bounded from above and below by the maximum and minimum acceleration of the car. 

We will now proceed with the simplest case of the OV model, where the optimal velocity function only depends on a single headway and is given by
\small
\begin{eqnarray}
V\left(h_n\right)=V_1+V_2\tanh\left(s_n\right), s_n=C_1(h_n-l)-C_2\label{vc}
\end{eqnarray}
\normalsize
The physical significance of different parameters in Eq.(\ref{vc}) can be found in\cite{bando,Helbing_PRE98}. We can now rewrite Eq.(\ref{general}) as
\begin{eqnarray}\label{ovmm}
\ddot s_n+\kappa_1\dot s_n=\kappa_2\left(\tanh s_{n+1}-\tanh s_n\right)
\end{eqnarray}
where $\kappa_1=\tau^{-1},\kappa_2=\tau^{-1}C_1V_2$. By rescaling the time variable $t\rightarrow \kappa_2t/\kappa_1$, the only dimensionless parameter in Eq.(\ref{ovmm}) is $\kappa=\kappa_1^2/\kappa_2$. This equation in general describes an array of particles moving in a viscous media with anisotropic non-linear nearest neighbour interaction.

We will now focus on Eq.(\ref{ovmm}), where $s_n$ is dimensionless. The change of variable in Eq.(\ref{vc}) not only tells us seemingly different driving behaviors are actually equivalent within the model; it also makes the symmetry of ODE's in Eq.(\ref{ovmm}) explicit. While the physical headway $h_n$ has to be positive, there is no such constraint on $s_n$; one should note the average of $s_n$ over all cars is inversely proportional to the linear car density of the lane with a shift, according to Eq.(\ref{vc}). Thus Eq.(\ref{vc}-\ref{ovmm}) completely define the physical model at hand, and mathematically Eq.(\ref{ovmm}) alone is sufficient. 

We will first discuss the properties of the individual clusters appearing in the solutions of Eq.(\ref{ovmm}). While many of these properties are known, here we derive them in the most general way. We also present the relations of the coexistence curve with the emergent extremal headways from the non-linear dynamics, which are not reported before and useful for numerical analysis. Linear analysis leads to a stable phase of $s_n=s_0$ against small perturbation above the spinodal line (or the neutral stability line) given by 
\begin{eqnarray}\label{linear}
2\text{sech}^2 s_0=\kappa.
\end{eqnarray}
In the regime $|s_0|>s_{c1}=|\text{sech}^{-1}\sqrt{\kappa/2}|$, a small perturbation to a uniform headway $s_0$ with $s_n(t\rightarrow 0)=s_0+\delta s_n$ leads to $s_n(t\rightarrow\infty)=s_0$. Here we take $\sum_n\delta s_n=0$ for technical convenience. Thus a random small initial variation of the positions of the cars in a single lane would not lead to the development of clusters, or traffic jams, in this regime. Note Eq.(\ref{linear}) is only exact in the limit when the perturbation goes to zero; close to the spinodal line, the uniform headway configuration is metastable, a large enough perturbation will also lead to the formation of clusters\cite{kerner}.

\begin{figure}
  \centering
  \setbox1=\hbox{\includegraphics[height=7cm]{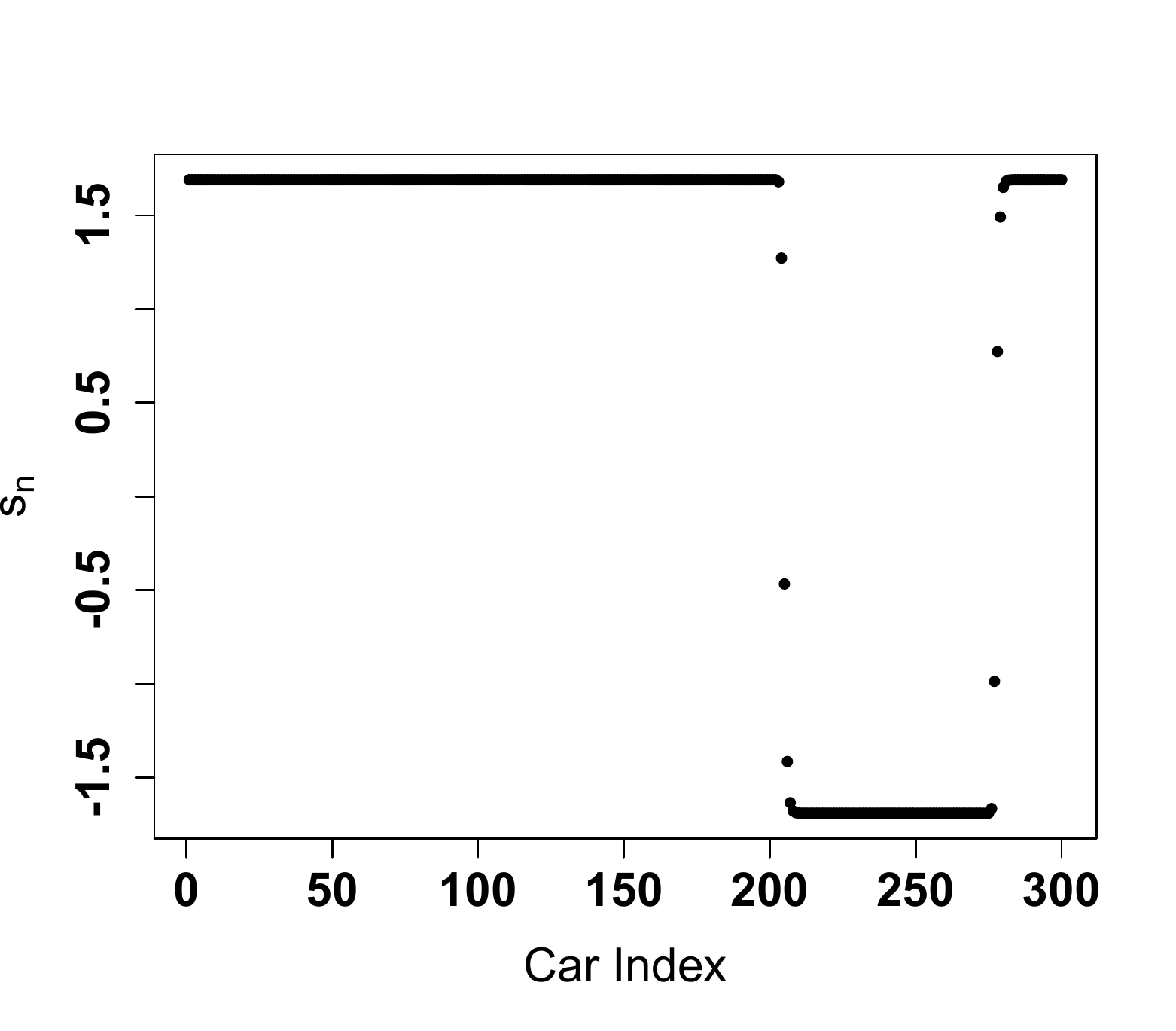}}
  \includegraphics[height=8cm]{single_peak.pdf}{\llap{\makebox[\wd1][l]{\raisebox{1.45cm}{\includegraphics[height=3cm]{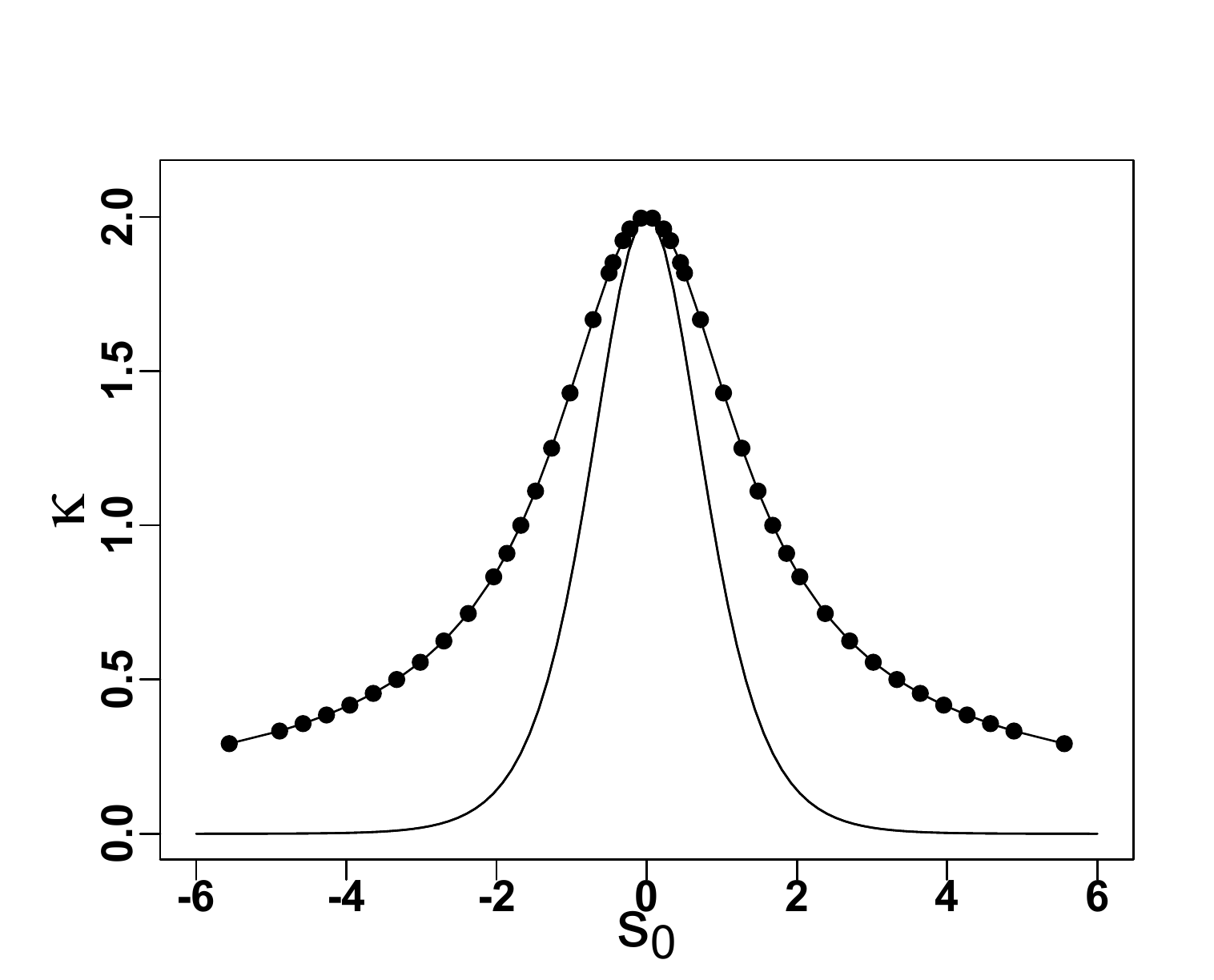}}}}}{\llap{\makebox[\wd1][l]{\raisebox{3.95cm}{\includegraphics[height=3cm]{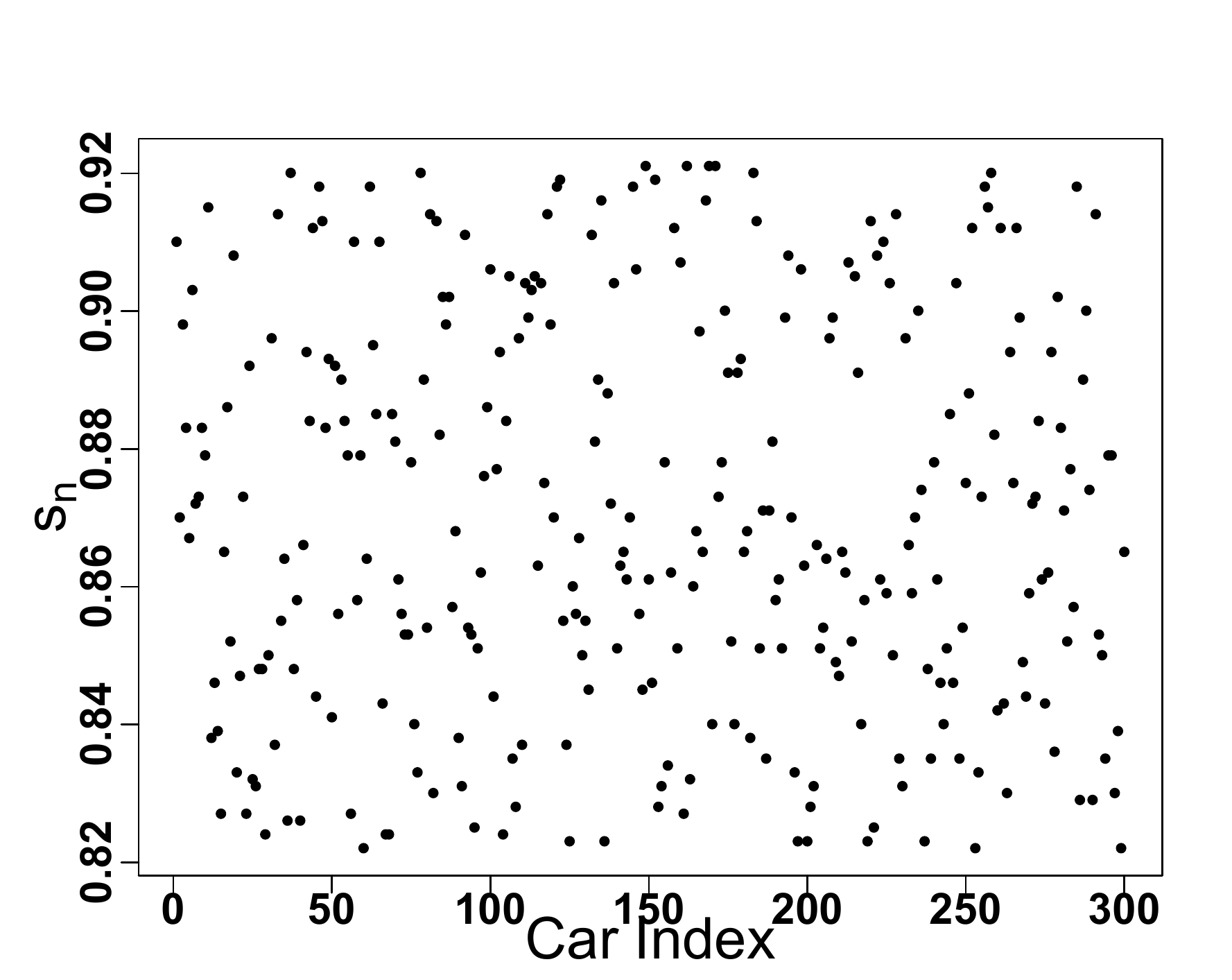}}}}}
  \caption{The plot of the headway as the function of the car index, when a jam or a cluster is formed. This cluster configuration evolves from a random initial headway distribution, as shown in the top inset. The bottom inset is the spinodal curve (the solid line without circles, plotted from Eq.(\ref{linear})), and the coexistence curve from numerical calculations (the solid line fitting the solid circles). The solid circles are numerically observed extremum headways at different $\kappa$.}
\label{singlepeak}
\end{figure}

We now show that the coexistence curve that separates the metastable phase and the absolutely stable phase can be numerically read off from the cluster formation alone. Firstly, in the regime $|s_0|<s_{c1}$, it is well known that small perturbations will grow in time with the formation of clusters, as shown in Fig.(\ref{singlepeak}), where a random initial condition settles into a configuration with the majority number of cars having two extremum headways given by $\pm s_{c2}$. As smaller $s_n$ implies higher physical car density, cars with headway $-s_{c2}$ form clusters or jams of very high density with minimal velocity, while cars with headway $s_{c2}$ moves with very high velocity, forming anti-clusters. Interestingly like $s_{c1}$, the numerical value of $s_{c2}$ only depends on $\kappa$ but \emph{not} on $s_0$, even for $s_0$ in the metastable regime. 

Secondly the number of cars involved in the ``kink" or ``anti-kinks" are independent of $s_0$ and the total number of cars $N$. A ``kink" is the ``go front", or the transition region from a cluster with $s_n\sim-s_{c2}$ to an anti-cluster with $s_n\sim s_{c2}$, while an "anti-kink" is the ``stop front", or the transition region from an anti-cluster to a cluster. Thus for large $N$  we can ignore cars in the ``(anti-)kink". Since the sum of the headways over all vehicles is conserved during the time evolution, the number of cars in the cluster is given by
\begin{eqnarray}\label{jnumber}
N_j=\frac{N}{2}\frac{s_{c2}-s_0}{s_{c2}}
\end{eqnarray}
Clearly for $s_0\ge s_{c2}$, no clusters can be formed, given random initial perturbations of any magnitude. Similarly, no anti-clusters can exist for $s_0<-s_{c2}$. We thus identify $s_{c2}$ as the coexistence curve \cite{kerner,yu,naka} and plot it together with $s_{c1}$ in Fig.(\ref{singlepeak}). The numerically calculated coexistence curve and the spinodal line coincides at the critical neutral stability point located at $s_0=0, \kappa=2$, agreeing with previous analysis\cite{nagatani,xue}. Note that $s_n$ can be negative, and the physical car density is calculated from Eq.(\ref{vc}). There is also a duality between $s_0\leftrightarrow -s_0$, where clusters at $s_0$ corresponds to anti-clusters at $-s_0$, and all behaviors at $s_0$ are identical to those at $-s_0$.

Progresses have been made in treating non-linear ODE describing car-following models analytically\cite{nakashini, yokokawa1,hayakawa,nagatani}; For Eq.(\ref{ovmm}) it is generally accepted that one can do a controled expansion near the critical neutral stability point and close to the neutral stability line; the former leads to modified KdV equations plus correction terms, that gives the approximate ``(anti-)kink" solutions; the latter reduces the original model to KdV equations plus corrections that give rise to soliton solutions\cite{kurtze}. However, away from the neutral stability line, it is clear from numerical calculation that if one makes the car index continuous, the transition between the two extremum headways is discontinuous and analytically intractable. 

One can, however, show that the ``kink" and ``anti-kink" of a single cluster move at the same velocity, by taking $s=\sum_{n=i}^js_n$. For the ``kink", the $i^{\text{th}}$ car is located in the cluster, while the $j^{\text{th}}$ car is located in the anti-cluster. From Eq.(\ref{ovmm}) we have
\begin{eqnarray}\label{calcv}
\ddot s+\kappa_1\dot s=2\kappa_2\tanh s_{c2}
\end{eqnarray}
The relevant set of solutions is $s=\left(2\kappa_2\tanh (s_{c2})/\kappa_1\right)t+C$, where $C$ is an unimportant constant of integration. This gives the velocity of the ``kink" as \emph{the number of cars per unit time} as follows
\begin{eqnarray}\label{kinkv}
v_{k}=\frac{\kappa_2}{\kappa_1}\frac{\tanh s_{c2}}{s_{c2}}
\end{eqnarray}
The velocity of the ``anti-kink" is calculated similarly, thus $v_k$ gives the velocity of the cluster, which again is \emph{independent} of the car density of the traffic lane. Here we make the assumption that for cars far away from the ``(anti-)kink", their headway takes the value of $\pm s_{c2}$. More importantly, if we concatenate two clusters together, as long as the assumption holds (e.g. when the two clusters are far away), they will move at the same velocity and will never merge.

\begin{figure}
  \centering
  \setbox1=\hbox{\includegraphics[height=3.9cm]{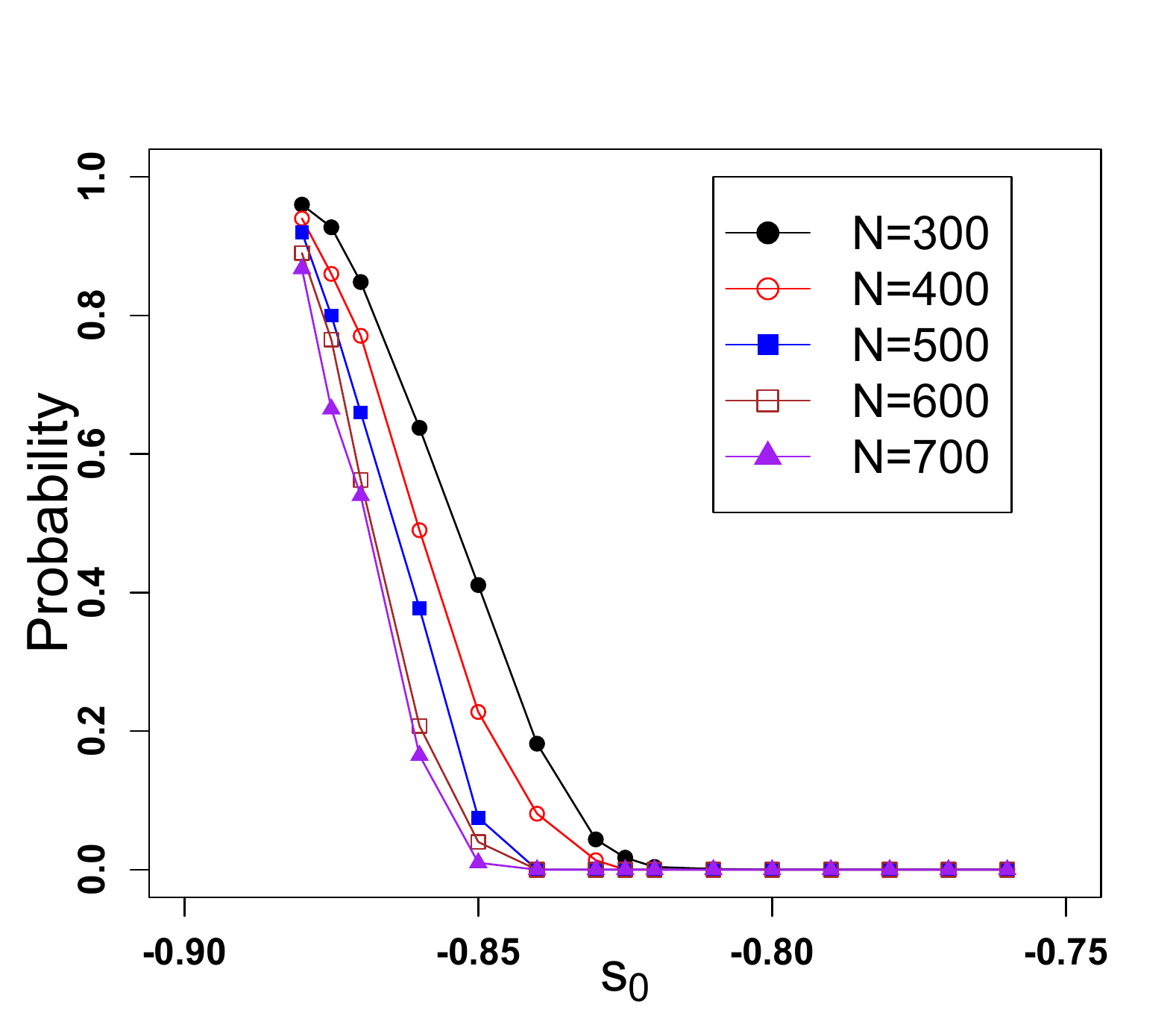}}
  \includegraphics[height=7.5cm]{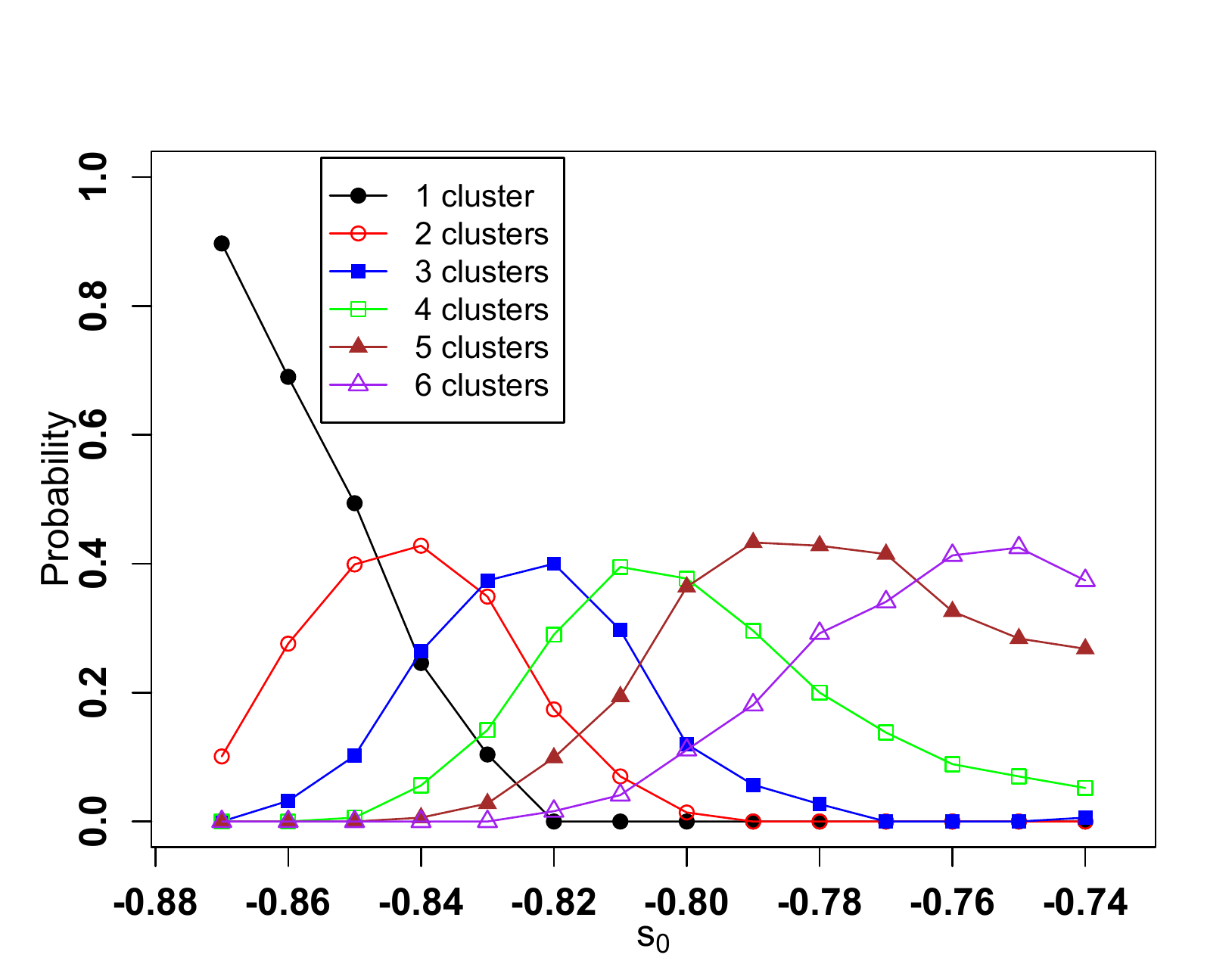}{\llap{\makebox[\wd1][l]{\raisebox{3.2cm}{\includegraphics[height=3.6cm]{probability2.pdf}}}}}
  \caption{(Color online) The probability of having one to six clusters in a single traffic lane, plotted as the function of the initial average headway $s_0$. The probability is calculated with three hundred cars and random initial headway perturbation. Inset: The probability of having only one cluster, as the function of the initial average headway. The probability is calculated for three hundred cars to seven hundred cars, showing some numerical evidence that in the limit of large number of cars, the probability curve converges to a well-defined limit. The probability is calculated at $t=30000 s$.}
\label{probability}
\end{figure}

We will proceed to study the dynamics of the multi-cluster solutions. One would naively expect that a random initial state like the inset of Fig.(\ref{singlepeak}) should lead to a random number of clusters\cite{zhang}, at least in the limit of large $N$, subjecting to the constraint of Eq.(\ref{jnumber}). However, our numerical results show that the probability distribution of the number of clusters is not random; it strongly depends on the initial headway $s_0$ and $\kappa$. We first calculate the probability distribution by fixing the strength of the initial random perturbation and $\kappa$ in Eq.(\ref{ovmm}), and only vary the initial headway $s_0$. For each value of $s_0$, sufficiently large number of random initial states are generated until the probability for each number of clusters converges. The probability distribution is plotted in Fig.(\ref{probability}), which is one of the main results of this work.

A few comments are in order here. In Fig.(\ref{probability}) we take $\kappa=1$ and only plot the part where $s_0$ is negative, because the probability distribution is \emph{identical} for $s_0$ and $-s_0$. For $|s_0|>0.87$ we can see the final state is dominated by one cluster, and this is true even for an infinitely long traffic lane as $N\rightarrow\infty$; in this case, most probably one very large cluster develops, instead of several clusters with smaller lengths. As $|s_0|$ decreases, the probability of having more than one cluster increases, and for $|s_0|<0.82$, it is almost impossible to have just one cluster. As $|s_0|$ further decreases towards \emph{zero}, the average number of clusters most probably will tend to infinity. This cannot be observed numerically for a finite number $N$, since at $s_0=0$ the total number of cars in the clusters is $\sim N/2$(see Eq.(\ref{jnumber})). For a physical traffic lane, from Eq.(\ref{vc}) the maximum number of jams will occur at car density $\sim \left(C_2/C_1+l\right)^{-1}$. Increasing or decreasing from that car density reduces the number of jams. This phenomenon of large number of ``phantom jams" occuring at some intermediate density could be used to empirically check the validity of the OV model.
 \begin{figure}
  \centering
  \setbox1=\hbox{\includegraphics[height=6.3cm]{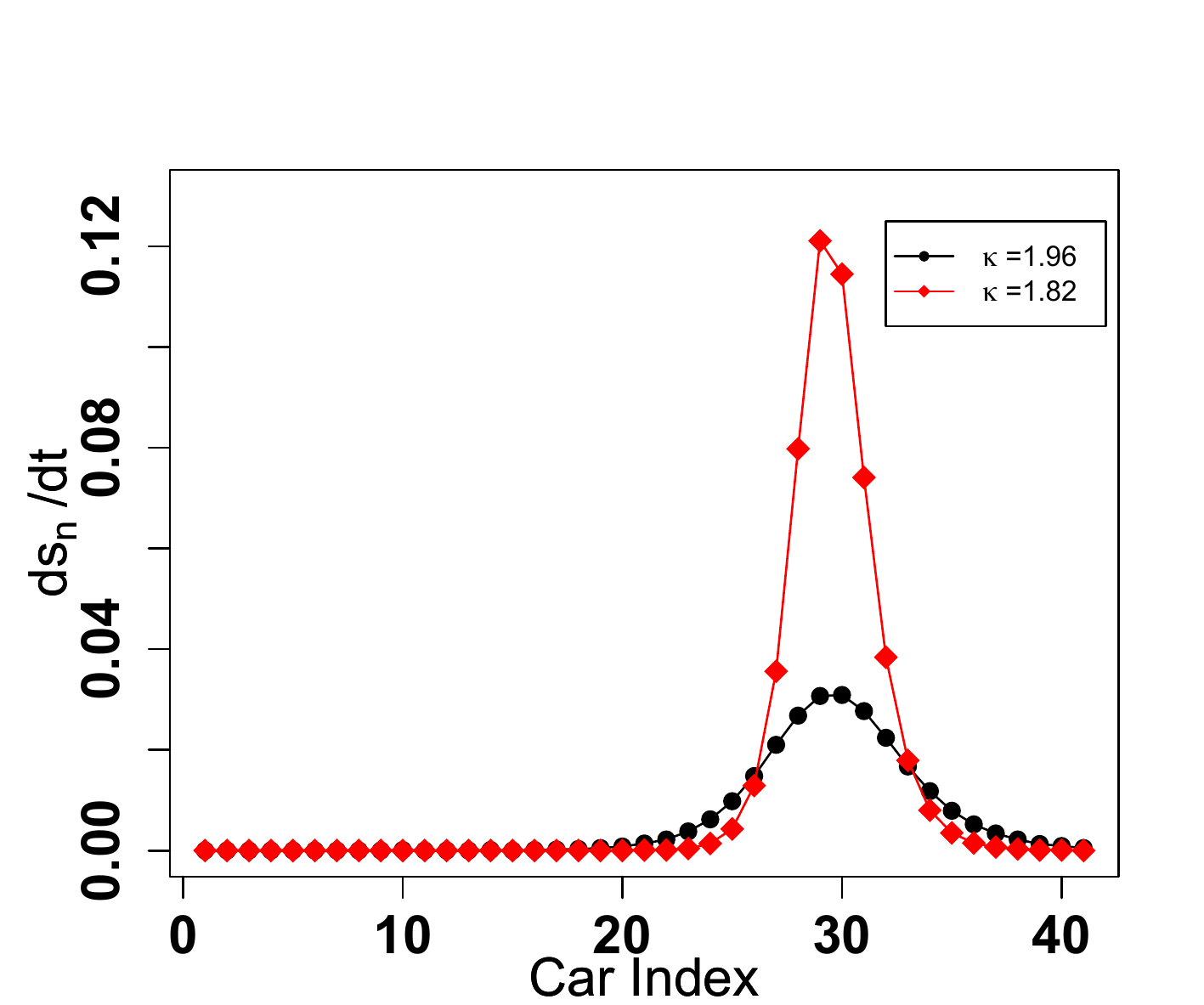}}
\includegraphics[width=9cm]{bump.pdf}{\llap{\makebox[\wd1][l]{\raisebox{1.2cm}{\includegraphics[height=3cm]{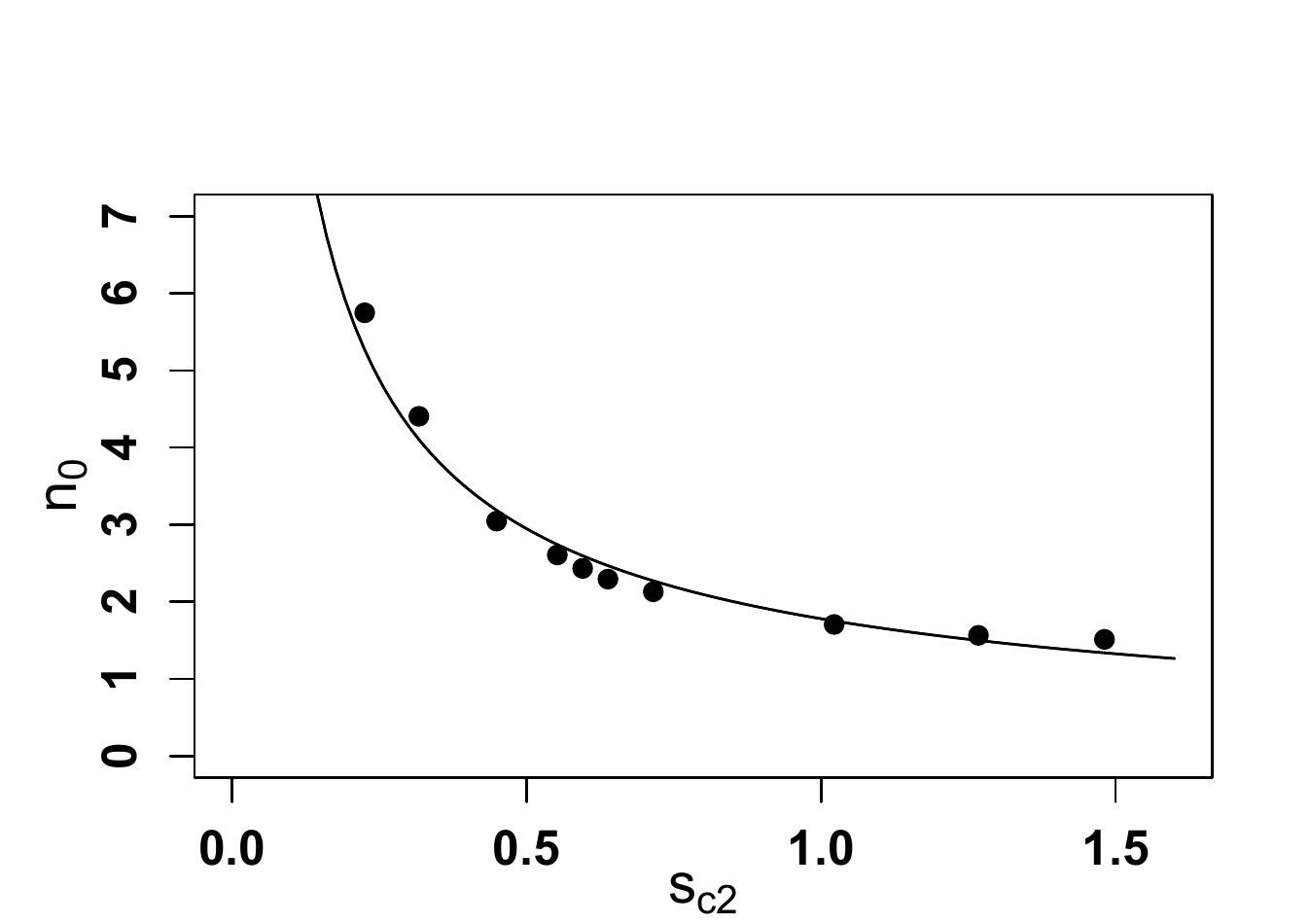}}}}}{\llap{\makebox[\wd1][l]{\raisebox{3.7cm}{\includegraphics[height=3cm]{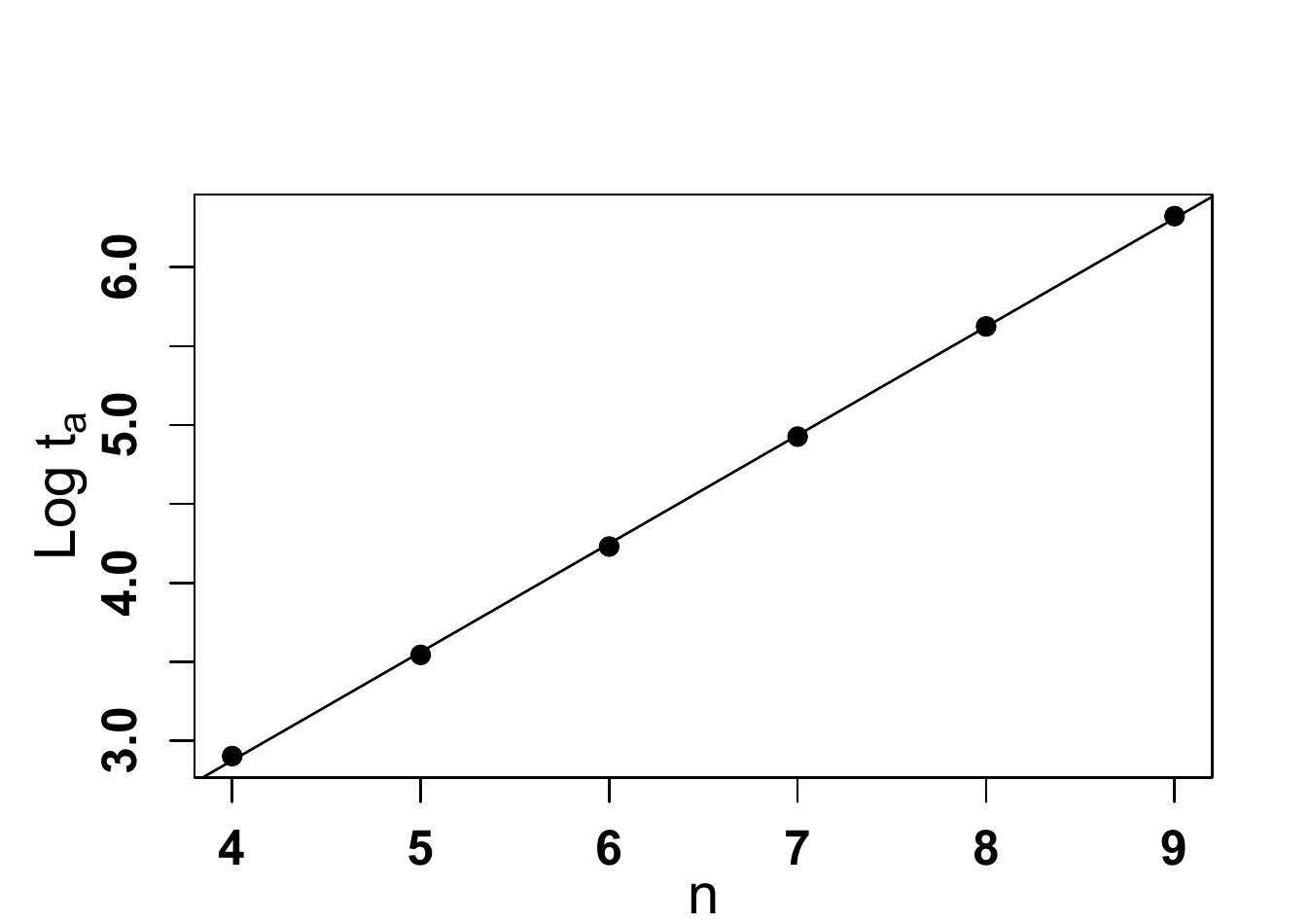}}}}}
\caption{The quasisoliton structure of $ds_n/dt$ as a function of the car index. The fewer the cars involved in the quasisoliton, the smaller the width of the quasisoliton, which depends only on $\kappa$ and not on the initial headway $s_0$. By convention a kink gives a positively charged quasisoliton as shown in this figure. An anti-kink gives a negatively charged quasisoliton. The top inset shows the dependence of the annihilation time $t_a$ on the number of cars between the quasisolitons of opposite charges, the exponential fit is numerically perfect. The intrinsic scale as a function of $s_{c2}$ is shown in the bottom inset.} 
\label{interaction}
\end{figure}

To understand the probability distribution of the number of clusters, we characterize quantitatively the strength of interaction between two clusters by the time it takes for them to merge. It is useful to plot $ds_n/dt$ instead of $s_n$ as a function of the car index $n$. The ``kinks" and ``anti-kinks" lead to exponentially localized ``quasisolitons" of opposite charges (see Fig.(\ref{interaction})), which closely resemble the ``autosolitons" in dissipative non-linear systems\cite{book}. When quasisolitons of opposite charges annihilate each other, two (anti-)clusters merge into one. We numerically observe that the time needed for annihilation, $t_a$,  increases exponentially with the number cars $n$ between the peaks of these two quasisolitons, giving the relationship
\begin{eqnarray}\label{ta}
t_a\sim e^{n/n_0}
\end{eqnarray}

While Eq.(\ref{kinkv}) dictates that kinks and anti-kinks travel at the same velocity, implying they would never merge, one should note the velocity is calculated from the extremal headways $s_{c2}$. The extremal headways are only attainable infinitely far away from the kink (or the anti-kink). Thus in principle, when multiple kinks and anti-kinks coexist in the same solution, they only move at the same velocity when they are infinitely apart. For finite separations, Eq.(\ref{kinkv}) is only an approximation, thus leading to the annihilations between the kink and anti-kink pair.
 
One thus note that when $|s_0|$ increases, the cluster (for $s_0>0$) or the anti-cluster (for $s_0<0$) region gets narrower(see Eq.(\ref{jnumber})), leading to higher probability of short distances between the quasisolitons. Thus the probability of having multiple (anti-) clusters is suppressed, as shown in Fig.(\ref{probability}). The intrinsic ``scale" $n_0$ in Eq.(\ref{ta}) depends on $s_{c2}$ or $\kappa$, which is also plotted in Fig.(\ref{interaction}). This is analogous to the interaction and collapsing of kinks and anti-kinks in the Ginzburg-Landau theory\cite{rougemont}, though here the total number of cars in the cluster has to satisfy Eq.(\ref{jnumber}), so that at least one cluster will remain for a finite system with periodic boundary condition. Thus the greater the intrinsic scale, the stronger the interactions between quasisolitons, so this scale can be used to quantify the absolute value of the quasisoliton charge. The interaction leads to merging of clusters, reducing the probability of having a large number of clusters in the traffic lane. 

\begin{figure}
  \centering
  \setbox1=\hbox{\includegraphics[height=5.37cm]{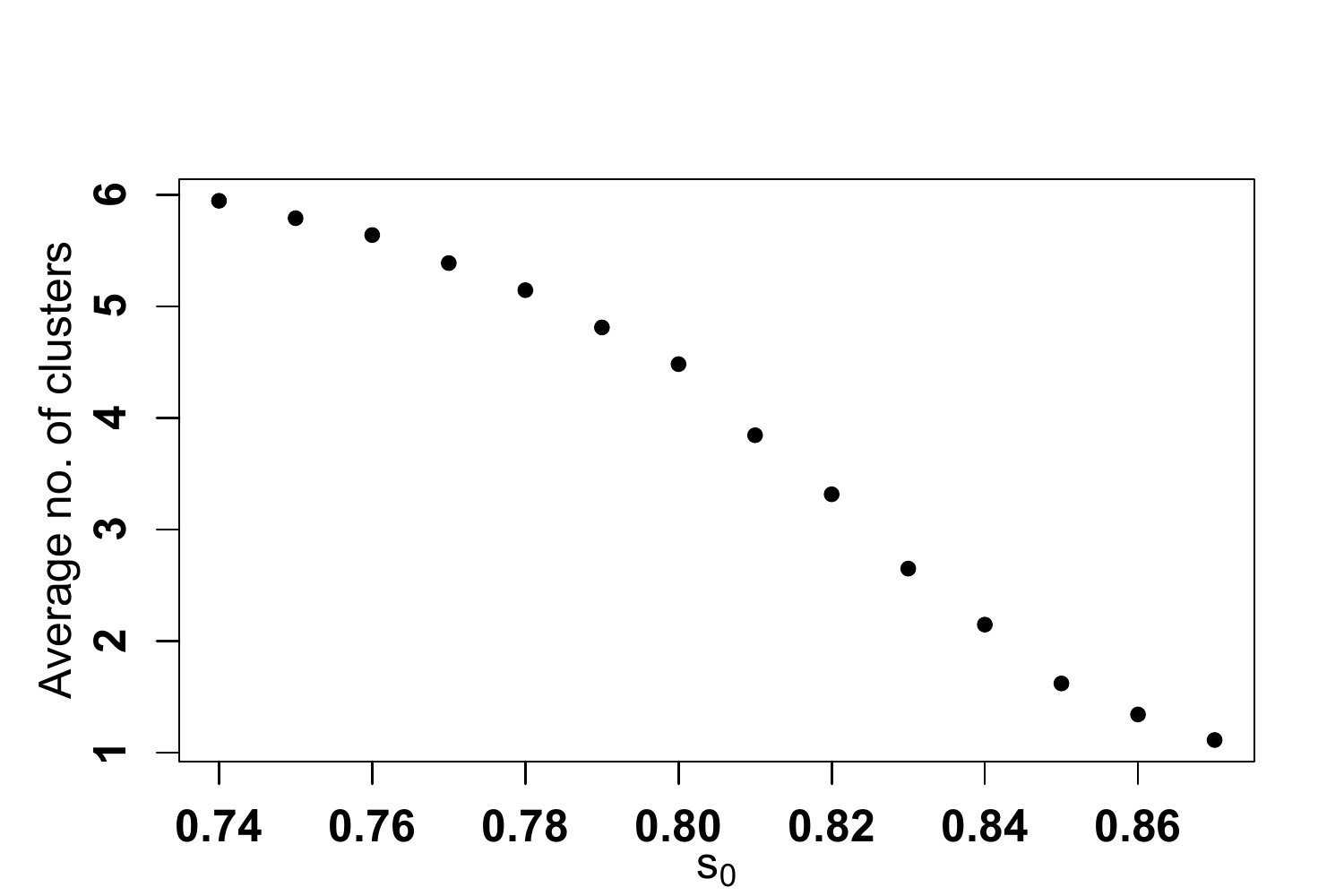}}
\includegraphics[width=8cm]{average1.pdf}{\llap{\makebox[\wd1][l]{\raisebox{-4.5cm}{\includegraphics[height=5.38cm]{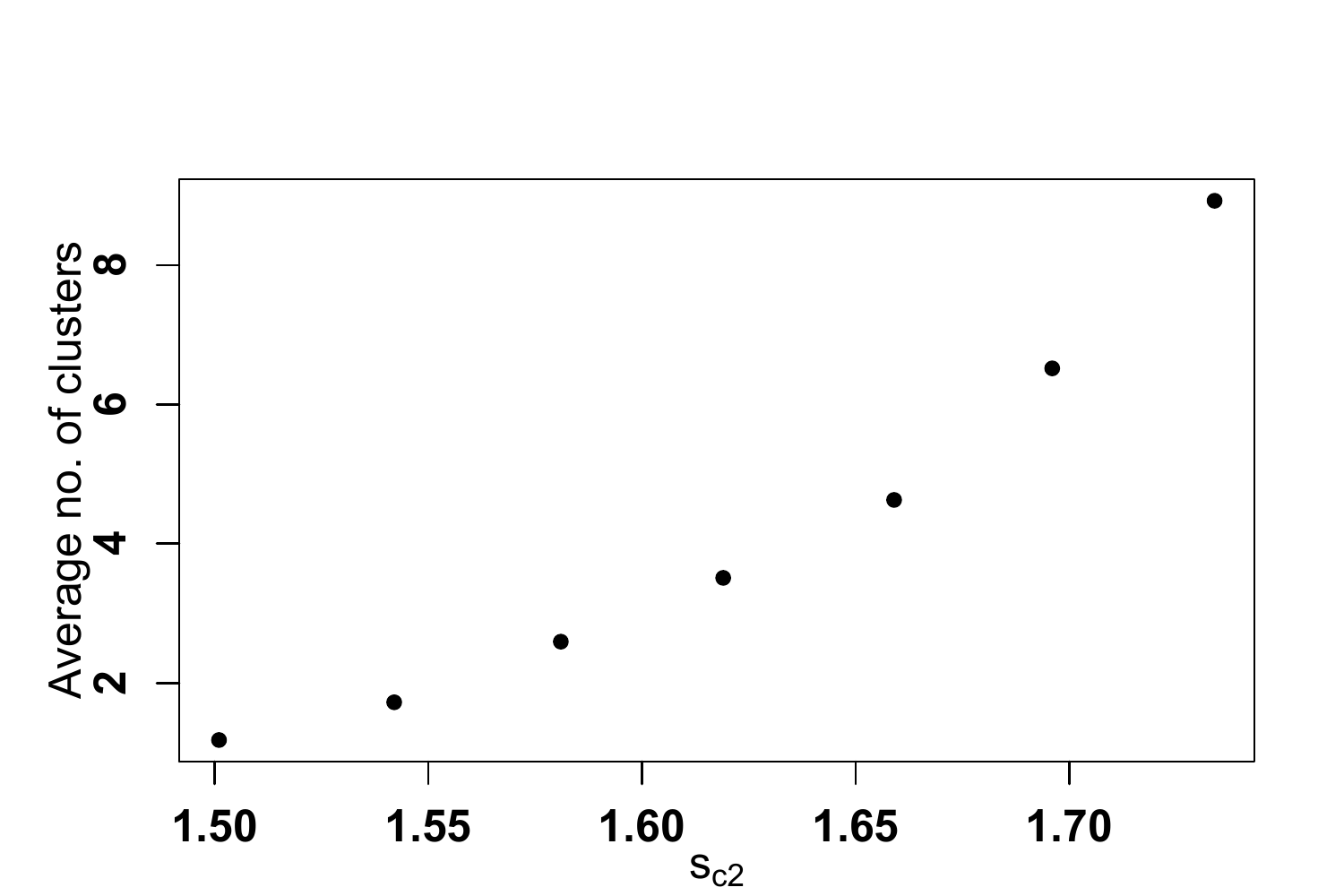}}}}}
\caption{The average number of clusters of a single lane traffic as a function of the intial headway (the top plot, while keeping the perturbation strength and $\kappa$ fixed), and as a function of $s_{c2}$ (the bottom plot, while keeping the perturbation strength and $s_0$ fixed).} 
\label{average}
\end{figure}

While the magnitude of the charge does \emph{not} depend on $s_0$, Fig.(\ref{probability}) will look qualitatively the same if the x-axis is replaced with increasing $s_{c2}$. The dependence of average number of clusters as a function of $s_0$ and $s_{c2}$ are plotted separately in Fig.(\ref{average}), numerically supporting the above explanation. For any finite number of cars, all clusters will eventually merge in the limit of very long time; thus the statements here are only rigorous in the limit that the number of cars $N\rightarrow \infty$. However because of the exponential dependence of the annihilation time on the number of cars between quasisolitons of opposite charges, the statements here are true for all practical purposes when the number of cars is reasonably large (even for computer simulation because of the finite numerical resolutions). 

We would also like to make a cautionary note here that both the cluster statistics in Fig.(\ref{probability}) and the average number of clusters in Fig.(\ref{average}) depend on the number of vehicles $N$ and the time of simulation $t$.  In principle, however, those two quantities are only well-defined in the limit of both $N$ and $t$ going to infinity. The finite scaling of the OV model is unfortunately very expensive numerically. On the other hand, the exponential dependence of the annihilation time on the number of cars between quasisolitons of opposite charges implies the cluster statistics and the average number of clusters converge very fast when $N$ increases (see also the inset of Fig.(\ref{probability})). One should also note that formally, the emergent quantities discussed in this paper are only well-defined when we take the limit of $N\rightarrow\infty$ first, followed by the limit of $t\rightarrow\infty$. 

In conclusion, we have investigated the OV model in the non-linear regime, where the metastable phase is delineated by the critical average initial headway $s_{c1}$ and $s_{c2}$. The behavior of the traffic jam evolution seems to be completely determined by the charge of, and the distance between, quasisolitons of opposite signs. This leads to non-trivial statistics of multiple clusters that depends both on $s_0$ and $s_{c2}$. This property is not only present in the OV model shown in details here. We have done extensive (but not necessarily thorough) numerical calculations for various extended OV models, which suggests that all features discussed above are qualitatively the same. A comprehensive and quantitative study of extended OV models will be presented elsewhere. Apart from its theoretical interest, we believe such studies are useful in designing and optimizing autonomous intelligent transport systems, where multiple clusters lead to undesirable wear-and-tear and need to be suppressed. It would also be interesting to see how the cluster statistics could be modified for more complicated traffic lanes with road works\cite{werner}. Given the universality of our results, it is also important to check the cluster statistics against the empirical data when modeling of real traffic dynamics is concerned, so as to understand what aspect of the real traffic complexity can really be captured by the General Motors model classes\cite{schreck}.

\begin{acknowledgements}
We would like to thank Prof. Weizhu Bao and Prof. Ren Weiqing from National University of Singapore for useful comments. This research was partially supported by Singapore A$^{\star}$STAR SERC ``Complex Systems" Research Programme grant 1224504056. The numerical calculations in this work is supported by ACRC of A$^{\star}$STAR. 
\end{acknowledgements}

\end{document}